# United Nations Digital Blue Helmets as a Starting Point for Cyber Peacekeeping

Nikolay Akatyev[1], Joshua I. James[2]
[1]Horangi, Singapore
[2]Legal Informatics and Forensic Science Institute,
Hallym University, Chuncheon, South Korea
nikolay.akatyev@gmail.com
joshua.i.james@hallym.ac.kr

**Abstract:** Prior works, such as the Tallinn manual on the international law applicable to cyber warfare, focus on the circumstances of cyber warfare. Many organizations are considering how to conduct cyber warfare, but few have discussed methods to reduce, or even prevent, cyber conflict. A recent series of publications started developing the framework of Cyber Peacekeeping (CPK) and its legal requirements. These works assessed the current state of organizations such as ITU IMPACT, NATO CCDCOE and Shanghai Cooperation Organization, and found that they did not satisfy requirements to effectively host CPK activities. An assessment of organizations currently working in the areas related to CPK found that the United Nations (UN) has mandates and organizational structures that appear to somewhat overlap the needs of CPK. However, the UN's current approach to Peacekeeping cannot be directly mapped to cyberspace. In this research we analyze the development of traditional Peacekeeping in the United Nations, and current initiatives in cyberspace. Specifically, we will compare the proposed CPK framework with the recent initiative of the United Nations named the 'Digital Blue Helmets' as well as with other projects in the UN which helps to predict and mitigate conflicts. Our goal is to find practical recommendations for the implementation of the CPK framework in the United Nations, and to examine how responsibilities defined in the CPK framework overlap with those of the 'Digital Blue Helmets' and the Global Pulse program.

## 1. Introduction

War is both destructive and constructive. Destruction is a threat that is responded to by passive and active protection; sometimes referred to as offensive and defensive security. Response to perceived threats tends to cause isolationism, or an us-vs-them mentality (Gladstein and Reilly, 1985). Much of the current discussion related to cyber warfare is focused on when and how to conduct aggressive activities, usually with the assumption of a national perspective.

The constructive aspect of war is based on threat avoidance. When groups wish to avoid the - usually costly - destructive side of war, they are more inclined to attempt peaceful cooperation. The challenge, however, is internal and external support for such peaceful cooperation. During physical conflict, one group may call a ceasefire and withdraw troops. This becomes an indicator of good faith in the peacemaking process. Acts such as these become a foundation upon which further peaceful cooperation can take place.

Cyberwarfare, however, attempts to utilize anonymity online as much as possible. With attribution nearly impossible, it is less obvious when good faith measures are taking place. In such a situation countries need more support for building peaceful cooperation in cyberspace. Support should come from impartial organizations with the goal of cyberwarfare prevention, cessation and response. This work examines proposed cyber peacekeeping models and their state of implementation.

To date, no organization has thoroughly addressed the roles and functions of cyber peacekeeping as defined by Akatyev and James (2015). Organizations such as ITU IMPACT, NATO CCDCOE and the Shanghai Cooperation Organization, among others, represent specific interests or address narrow problems. The United Nations Peacekeeping operations do not address challenges in cyberspace, and their functions cannot be directly mapped to cyberspace operations. Other branches of the United Nations (UN) are starting to consider the problem of peacekeeping in cyberspace, and it appears as though the UN may be one of the best options for hosting Cyber Peacekeeping efforts in the future. A recent peacekeeping development called the United Nations 'Digital Blue Helmets' program ("Digital Blue Helmets", n.d.) appears to focus on Dark Web and critical infrastructure issues. However, digital and physical conflict cessation and prevention demand much greater scope. Another UN program, the Global Pulse, focuses on data analysis and has experience with digital technologies. This program's structure and experience is useful for cyber peacekeeping support, but, again, lacks scope and focus on such issues.

## 1.1 Contribution

While past work defined organizational structure, functions, and legislative requirements of Cyber Peacekeeping (CPK), little consideration was given to the concrete implementation as a separate organization or within a structure of an international body. This work will analyze cyber peacekeeping frameworks and related organizations, with a specific focus on the United Nations Peacekeeping activities and cyber security initiatives. We then update prior framework recommendations based on current work from the international community, and propose implementation plans that use existing organizational efforts.

## 2. Cyber Peacekeeping

The main goal of Cyber Peacekeeping is to promote online safety and security that assists in both physical and cyber conflict cessation, and helps protect cyber civilians from becoming either victims or participants in cyber conflicts (Akatyev and James, 2015). Protection and prevention is provided through pre and post conflict monitoring, cleanup and capacity building, as well as response and coordination activities during conflicts.

Cyber Peacekeeping seeks to prevent and mitigate cyber and physical conflicts before the conflict escalates. Further, CPK works towards conflict cessation during periods of conflict. These goals are achieved through cyber conflict prevention, mitigation, post-conflict containment and rehabilitation services. Two specific implementations of mitigation services previously proposed include the concept of a Cyberspace Safe Layer (CSL), and an Information Clearinghouse (ICH). The CSL addresses the need to define and protect critical cyber infrastructure and help delineate unethical targets in conflicts (Schmitt, 2013). The ICH helps in the tempering of rumor and bias on social networks that is likely to lead to the escalation of digital and/or physical conflicts, and potential recruitment of unaffiliated actors (Nissen, 2016; "Digital Blue Helmets", n.d.).

Cyber Peacekeeping differs from UN Peacekeeping in scope. Field (1993) claims that UN Peacekeeping is different from peacemaking and peacebuilding in both utilized means and targeting goals. The goal of UNPK strives to prevent the recurrence of violence. UNPK may actively apply force when it overlaps with peacemaking. They may also start building on the safety and stability of a state assuming some peacebuilding roles, but these are not the main activities of UNPK. In cyberspace violence can spread fast, and peacekeeping efforts should also try to prevent conflict escalation before conflict occurs.

## 2.1 The Cyber Peacekeeping framework

To carry out its mission, we define goals, roles and functions for Cyber Peacekeeping as shown in Figure 1. Each role of Cyber Peacekeeping can contribute to the safety and security of cyberspace at all three stages of a conflict: no conflict, during conflict, after conflict. For example CPK as a guardian will monitor potential threats when there is no conflict. During conflict CPK will stop the spread of cyber attacks and involved cyber weapons responding with defensive counterattacks as a last resort for "self-defense or defense of the mandate" ("Principles of UN peacekeeping", n.d.). After conflict CPK as a guardian will lead cleanup activities related to distribution and alteration of cyber weapons. Figure 1 shows relations among roles and their functions for different stages of a conflict depicted by different types of lines: solid (guardian), dot (mediator), dash (coordinator), dash-dot (builder).

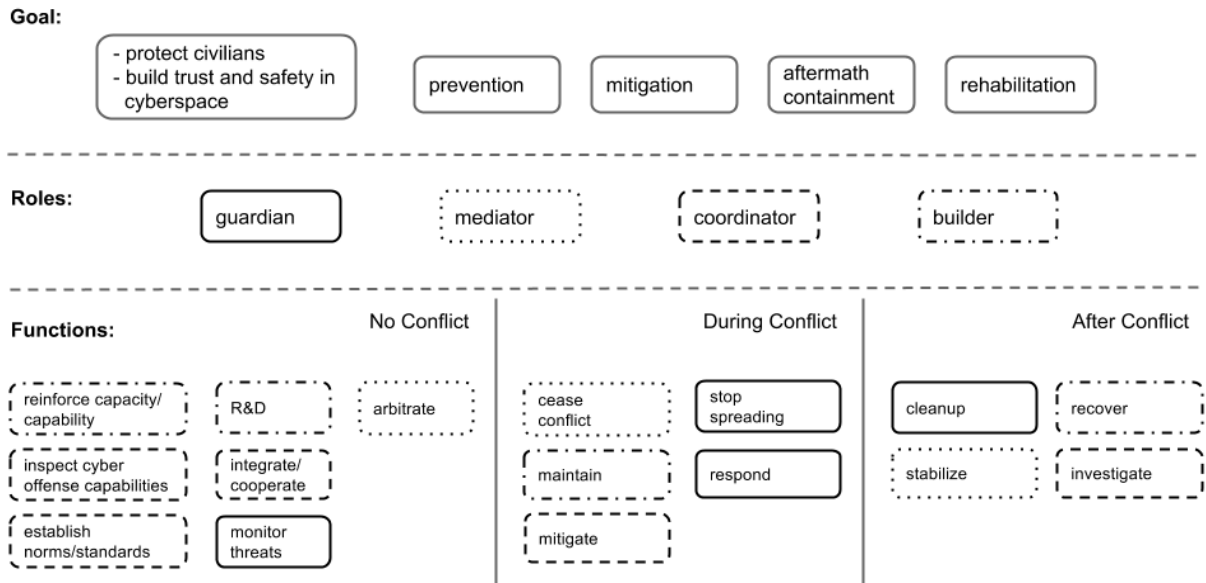

**Figure 1:** Overview of the framework of CPK reflecting layers of goals, roles and functions when there is no conflict, during conflict and after conflict. Solid line, guardian role and related functions; dotted line, mediator and related functions; dashed line, coordinator and related functions; dash-dotted line, builder and related functions

The goals of Cyber Peacekeeping are defined as:
- Protect civilians
  - The main goal of CPK is the protection of civilians.  CPK must be impartial to any State independent of contributions.
- Increase trust and security in cyberspace
  - Through conflict prevention, mitigation and rehabilitation tasks, trust in cyberspace can be maintained and security increased.
- Prevention
  - Focuses on preparation for potential attacks, and preventing cyber conflict escalation when conflicts begin
- Mitigation
  - Focuses on containing conflicts and minimizing damage to infrastructure and civilians
- Aftermath Containment
  - Focuses on containment of tools and information that may be re-purposed or reused in other conflicts, as well as using collected information for prevention
- Rehabilitation
  - Focuses on rebuilding infrastructure, security and trust post-conflict

## 2.2 Cyber Peacekeeping Requirements

Prior works (Akatyev and James, 2016) looked at the legislative requirements for organizations to begin cyber peacekeeping efforts. The formal requirements for establishing Cyber Peacekeeping are minimal but require considerable political action:

- CPK does not have operational authority within member states, and does not need additional legal frameworks for any of its main functions
- Future development of a legally-binding collaboration will be necessary for CPK to be effective as threats of cyber conflicts grow
- To support Cyber Peacekeeping functions, member states need to formally accept terms, definitions and concepts that CPK works in, including differentiating terms such as 'freedom of speech', 'propaganda' and 'inflammatory information'
- It also includes a formal agreement on the protection of civilians through the protection of critical infrastructure
- States need to determine how the UN Charter exactly maps to different activities in cyberspace

- A formal agreement would require specifying the CPK governance structure, and especially the terms by which CPK could operate in conflict and non-conflict situations in cyberspace

While formal requirements mostly relate to organization and oversight, informal requirements deal more with the practice of Cyber Peacekeeping. Specifically, the establishment, promotion (and potentially enforcement) of cyber norms. These cyber norms will be the basis for CPK operations.

An essential part of the establishment of cyber norms is confidence building. CPK would work with States and global and regional organizations in order to establish understanding among different groups with their own values, and to promote best practices of appropriate behaviour in cyberspace. The culture of training and information sharing already exists to some extent. Some groups also run anti-cyber terrorists operations and share technologies (Goldman, 2016; O'Connell, 2016). CPK will endorse these activities and facilitate their globalization to meet the goal of digital and traditional conflict prevention, mitigation and cessation.

After cyber norms (and formal agreement of terms), informal agreements will need to be established between CPK and individual member countries regarding services and access. For example, one function of CPK is to help secure, support and maintain critical infrastructure. First the scope of critical infrastructure would need to be formally defined, and informal agreements would need to be made with each member country regarding how CPK would help protect their infrastructure. Some members may allow full access to ensure CPK can properly maintain and monitor security, where other members would prefer CPK to be only an outside monitoring organization.

Informal requirements mostly deal with CPK member country agreements and permission to interact internally and externally to the member country. The conditions for that support, and the requirements member countries must meet to continue to receive such support from the CPK.

### 2.3 Current status of CPK-related organizations

Muller (2016) explains the failure of the multi-stakeholder approach with examples from the Internet Governance Forum (IGF) and the World Group on Internet Governance (WGIG). The core problem of the multi-stakeholder approach's implementation includes separation of public collaboration from private collaboration. Specifically, governmental multi-stakeholders organize their own group as well as private organizations and NGOs develop their own, separate initiatives. Instead of supporting the functions of each other, public and private sectors collaborate through a rigid interface which is difficult to use and develop. For example, the failed "Alert System for Digital Infrastructure" (VDI) sensor program in Norway.

Another example of the public-private cooperation was ITU IMPACT ("IMPACT", 2015) which formally become the cybersecurity executing arm of the United Nations' specialised agency; the International Telecommunication Union (ITU). ITU IMPACT differentiated itself by bringing together multiple countries as well as prominent private organizations. However, multiple factors such as no involvement of major national cyber powers, and a focus of the organization on training and monitoring for businesses, limited the scope of ITU IMPACT's responsibilities.

Regional organizations such as NATO and the Shanghai Cooperation Organization recognized the significance of threats in cyberspace. After cyber-attacks on Estonia, NATO included cyber threats into Article 5 (NATO, 2014). The SCO discussed cooperative protection in cyberspace among participating members, and proposed a supervision role for the UN. However, activities of regional organizations were limited to their member-states, and focused mostly on military components. Without global confidence-building and understanding, these regional initiatives may only escalate conflict.

## 3. The United Nations Peacekeeping efforts

Peacekeeping was first envisioned by the drafters of the UN Charter in 1945, and the first operation was conducted in 1948 (Field, 1993; Bellamy, et al., 2010). Despite their plan for the creation of a large standing military force, such a force was not possible during the Cold War. Later, with the support of the Canadian initiatives, UN-sanctioned peacekeeping missions were made a reality.

For more than sixty years UN peacekeeping (UNPK) operations played different roles, and though there has been much criticism for a series of failures, there are claims that the benefits of peacekeeping outweigh the challenges (Fortina, 2008). UNPK evolved significantly from narrowly-purposed missions into multi-functional campaigns partially covering peacemaking and peacebuilding tasks in some situations (Bellamy, et al., 2010).

In the current information era, different divisions in the UN have started paying attention to cyberspace. The Global Pulse program ("United Nations Global Pulse", 2016) attempts to apply large-scale data analysis to humanitarian challenges, helping to predict humanitarian catastrophes such as diseases, environmental problems or civil aggressions in unstable countries. Karlsrud (2014) analyzes this program and how it can be applied to the needs of peacekeeping. He elaborates on specific cases of the use of the data for the facilitation of the current types of UNPK operations.

The most recent UN initiative by the UN Office of Information and Communication Technologies (OICT) proposes the Digital Blue Helmets (DBH). The DBH is a first step in shifting focus to problems in cyberspace, like Dark Web issues. However, the main function of DBH is the protection of UN cyber infrastructure itself ("Digital Blue Helmets", n.d.).

### 3.1 United Nations Mandates

Through the execution of UNPK operations, the Security Council (SC) satisfies its primary responsibility from the UN Charter for the maintenance of international peace and security ("Mandates", n.d.). UNPK operations are deployed on the basis of mandates from the UN SC. International human rights law described in The Universal Declaration of Human Rights is a significant part of the normative framework for UNPK operations.

Responsibilities of UNPK vary from mission to mission, but common peacekeeping tasks include prevention of the outbreak and spill-over of conflict, stabilization after the conflict and peacebuilding tasks. Peacebuilding covers the protection and promotion of human rights as well as disarmament and demobilization.

In *Principles of UN peacekeeping* (n.d.) the organization tracked how UNPK developed beyond monitoring ceasefires to multidimensional peacekeeping operations. This changed followed due to the reality of new intra-state conflicts that mostly substituted inter-state conflict. However, inter-state conflicts are currently moving into cyberspace. It appears as though the goals and functions of Cyber Peacekeeping, as previously defined, fall under United Nations Peacekeeping and Human Rights mandates.

### 3.2 UN Peacekeeping Operations

Current UNPK operations are beginning to use more technology, such as cell phone movement data to determine challenging areas, and drones to quickly assess dangerous areas (Karlsrud, 2016). However, UNPK is limited in data, tools analysis and human resources to process such information. UNPK is focused on physical conflict areas, and so far has not shown an interest in incorporating cyberspace - such as activities in social media - into their peacekeeping efforts. Cyberspace can greatly affect physical conflict (Loukas, 2015). Until now UNPK has shown an interest in data related tools to assist in physical peacekeeping without considering that physical conflict can potentially be avoided altogether through peacekeeping in cyberspace.

### 3.3 Digital Blue Helmets

Digital Blue Helmets is a recent initiative, and comprehensive strategy does not yet exist for the organization. Information on the official website ("Digital Blue Helmets", n.d.) and an interview with Atefeg Riazi, UN CITO from the Office of Information and Communication Technologies, (Tucci, 2016) shows similar values and vision between the DBH and CPK. The DBH is a program under the OICT in the United Nations, and acts under United Nations mandates.

In the long term, the DBH will focus on cybercrime, and specifically cyber-underground marketplaces. Sources discuss the risk of the escalation of conflicts as the result of "the power of social media to attract new recruits" ("Cyber Risk", n.d.), as well as cyber terrorism and cyber threats to the critical infrastructure. The DBH already considers these issues as a challenge of national boundaries and international law. This challenge is innate in such an organization, and is one of the limitations of the United Nations as a host for CPK activities.

As the result of the challenges of the boundaries and international law in the cyberspace, the DBH aims to involve "all stakeholders, including the United Nations Secretariat, Agencies, Funds and Member States, as well as external partners, including academia, the public and private sectors and the [general] public" ("Digital Blue Helmets", n.d.). Building the DBH as a group of experts means a permanent but less consistent contingent, but clarification is still needed as to how they would be recruited, either as professionals or contributed by member states, as well as what skills they would possess. Conducting only research and sharing information would bring them to the territory of ITU IMPACT, so the DBH needs offensive and defensive cybersecurity skills to differentiate themselves.

However, the UN CITO declared the mission of the DBH to "operate in the cyber world protecting the UN from cyber intrusion" (Tucci, 2016). This seems to indicate a focus on UN infrastructure, but not necessarily defending the interests of world peace.

### 3.4 Global Pulse
The United Nations formed Global Pulse (GP) to promote Big Data applied to development and humanitarian action.

> *[Global Pulse's] vision is a future in which big data is harnessed safely and responsibly as a public good. Its mission is to accelerate discovery, development and scaled adoption of big data innovation for sustainable development and humanitarian action… Global Pulse is working to promote awareness of the opportunities Big Data presents for sustainable development and humanitarian action, forge public-private data sharing partnerships, generate high-impact analytical tools and approaches through its network of Pulse Labs, and drive broad adoption of useful innovations across the UN System ("Global Pulse", n.d.)*.

Global Pulse is applying data analysis to a number of public data sources including social media. The DBH and GP have the potential to converge in their data collection and analysis tasks, and include cyber threat detection, prevention and mitigation. This partnership does not currently appear to be developed between the two organizations.

Karlsrud (2014) analyzed the GP and its application to peacekeeping operations. He noted how new technology is used by groups spreading disinformation with the purpose to incite hatred. It is overlapping areas like these where GP and DBH could work together.

The GP is a useful initiative for the incorporation of information technologies in the UN, and can be extended for peacekeeping activities. However, the GP is currently focusing on solutions to humanitarian problems, and it is unclear if an expansion of scope into cybercrime and cyber warfare would be covered by the mandate of GP, which is currently focused on innovation and data sharing. Karlsrud (2014) examined how the GP could be applied for the UNPK, which is limited to the application of the GP as a tool for the physical operations. The GP may in time help prevent violent physical conflicts, but the application of big data can also be useful for the prediction and prevention of both physical and cyber conflicts. The GP can be a tool for both the UNPK and CPK as it can assist but does not by itself solve problems of violence, conflict escalation and threats to peace in cyberspace.

## 4. Comparison of CPK, GP and DBH

The DBH and CPK share a similar vision and have similar approaches to recognized cyber threats. However, the DBH has yet to establish a concrete plan for the implementation of its vision. CPK already has a proposed organizational and activity framework developed that has been analyzed in terms of its practical implementation. CPK's initial scope of activities has been defined to support conflict avoidance, cessation and recovery. For all the framework development and research on CPK, it still lacks the backing of an organization that can work at the scale needed for global cyber peacekeeping. We propose the DBH start working to implement the CPK framework, and specifically on 'low hanging fruit' such as the CPK-proposed Cyberspace Safe Layer and Information Clearinghouse (Akatyev and James, 2015). After reviewing UN mandates, it appears that the objectives of the Information Clearinghouse and Cyberspace Safe Layer define similar goals to the UN CITO's vision for the "light web" (Tucci, 2016), a web space designed to counter the negative effects of the dark web.

The experience of the GP could also greatly contribute to the development of the combined DBH and CPK organization by applying its developed expertise in data collection and analysis from social media and other sources. Specifically, this expertise would greatly enhance capabilities of the Information Clearinghouse to combat issues such as "fake news" and government or other organization-sponsored misinformation campaigns.

### 4.1 Digital Blue Helmets as a Starting Point for Cyber Peacekeeping

As the Canadian push for peacekeeing played a pivotal moment in the implementation of the UNPK operations (Field, 1993), the UN CITO's DBH initiative can be a starting point for the development and implementation of CPK. In section 3, we identified that the plan for peacekeeping, as initially drafted, included a permanent, global peace force. Both politics and resources made a permanent peace force impossible. Location, resources, security and training make peacekeeping efforts difficult for peacekeepers, and any type of participation - except donations -  at an individual level is nearly impossible.

Similar to physical conflict, conflict in cyberspace has a potential to escalate extremely quickly in more scenarios. This could result in retaliation in cyberspace, or spill over to exacerbate physical conflict. Peacekeeping, however, is not only about responding to conflict, but helping to prevent conflict before it begins. For this reason, peacekeeping activities need to be taking place constantly. Many organizations and individuals need to take ownership of the peacekeeping process, similar to organizations such as Security Without Borders ("Security Without Borders", n.d.), where individuals with different skillsets volunteer to secure NGOs and other non-profits from spying and exploitation. The nature of cyberspace allows this kind of global collaboration to happen in near-real-time.

As discussed, we recommend that the DBH, with the support of GP and the UNPK, begin to implement the CPK framework as an expansion of their current work. Being under the UN umbrella, the DBH can benefit from the UN's international cooperation experience and especially from lessons learned from the UNPK. At the same time, the DBH should seek to use the benefits of globally connected experts with similar goals. Rather than keeping complete, centralized control, DBH should become a type of CPK coordination center. Where the goals and scope of CPK are concretely defined and developed to provide guidance to semi-independent cyber peacekeeping groups around the world. This distributed network would allow a 24-7 peacekeeping force for fast response and intelligence gathering. In the reality of the multi-stakeholder globally connected cyberspace, the UN needs to create a permanent cyber peacekeeping force that can rapidly, globally address cyber threats.

Karlsrud (2014) separated conflict prevention and humanitarian actions from peacekeeping and noted that the GP only focused on the first two activities. However, the scope of CPK includes all activities that can impact the peace in the cyberspace. We can not ignore conflict prevention or humanitarian actions because cyberspace is interconnected and global, and military cyber attacks targeting state assets can easily affect civilians. Moreover, conflicts in cyberspace spread so fast and broad that conflict prevention is potentially more difficult but more important than in the physical world. Hence, the experience of the GP in conflict prevention and humanitarian actions are indispensable for the CPK.

Prior CPK research showed that the development of such a politically sensitive organization should be started with small, specific steps (Akatyev and James, 2015). It is important to have a clear vision, goals and outlined full-fledged framework that is supported by major stakeholders.

### 4.2 Multi-stakeholder foundation

As discussed, the CPK is potentially implemented on the foundation of UN initiatives; DPH, GP, UNPK and others. This effort would also require guidance from other cybercrime and cyber-conflict related organizations which are included in this research. INTERPOL has an established structure and constitution; ITU IMPACT provides an example and lessons learned for global cooperation in the cyberspace; NATO CCDCOE has experience in the defence in cyber-conflicts; and Shanghai Cooperation Organization (SCO) proposes concrete roles for the leadership of the UN in the cyberspace security activities.

Ramjoue (2011) claims that "UN member states have been wary of UN peacekeeping operations gaining intelligence-gathering and analysis capabilities". Karlsrud (2014) believes that this attitude is shifting as the UN realizes that intelligence gathering can have a profound impact on the success of peacekeeping activities. This is somewhat confirmed with the GP, and the analysis of big data, which has the potential to extract a considerable amount of actionable intelligence for humanitarian and peacekeeping purposes. The same is true for cyberspace. Intelligence gathering and analysis is critical to make strategy decisions towards conflict avoidance in cyberspace. This data gathering, however, cannot be done alone. No single organization has access to all relevant data sources and the ability to extract and digest relevant intelligence. The DBH/CPK then needs to utilize many stakeholders. Partners should include the UN's own branches, private business, security research companies, the financial sector, academia, hackers, security experts and cyber underground experts. Such collaboration, however, is not a usual structure for the United Nations.

### 5. Conclusions

Cyberwarfare is happening. Attribution makes the differentiation of cybercrime and cyberwarfare difficult, but there are cases around the world with attacks that appear to be government sponsored. Beyond a growing threat to cyberspace, cyber conflict can - and has - spilled over to affect physical conflict. UNPK seeks to prevent the reoccurrence of conflict. Cyber Peacekeeping needs to do even more; focusing not only on conflict cessation, but also pre-conflict prevention and post-conflict management. This work described the framework and requirements for Cyber Peacekeeping, and analyzed potential organizations that could support the vision and scope of CPK. We believe one of the best organizations to support CPK is the recently created UN Digital Blue Helmets program. Currently Digital Blue Helmets is limited in scope and definition, specifically focusing on dark web issues and protection of UN infrastructure. Despite the limited scope, the DBH appears to be a strong base for the implementation of CPK. DBH with the support of GP and UNPK would have the information, analysis and experience to implement the first stages of CPK. By focusing on community building and inclusion of a distributed network of experts around the world, DBH/CPK could influence peacebuilding in both physical and digital spaces.

**Acknowledgements**

This research was supported by Hallym University Research Fund, 2016 (HRF201603007).